\makeatletter
\@ifundefined{@parse@version@dash}{%
\def\@parse@version#1{\@parse@version@0#1}
\def\@parse@version@#1/#2/#3#4#5\@nil{%
\@parse@version@dash#1-#2-#3#4\@nil}
\def\@parse@version@dash#1-#2-#3#4#5\@nil{%
  \if\relax#2\relax\else#1\fi#2#3#4 }
}{}
\makeatother
\documentclass[reprint,showkeys,longbibliography,superscriptaddress,aps,amsmath,amssymb,notitlepage,prl]{revtex4-2}
\usepackage[dvipsnames]{xcolor}

\usepackage{graphicx}
\usepackage{dcolumn}
\usepackage{bm}
\usepackage{multirow}
\allowdisplaybreaks
\usepackage{verbatim}
\usepackage{xspace} 
\usepackage{float}
\usepackage{booktabs}
\usepackage[colorlinks]{hyperref}

\newcommand\myshade{80}
\colorlet{mylinkcolor}{ForestGreen}
\colorlet{mycitecolor}{Red}
\colorlet{myurlcolor}{violet}

\hypersetup{
  linkcolor  = mylinkcolor!\myshade!black,
  citecolor  = mycitecolor!\myshade!black,
  urlcolor   = myurlcolor!\myshade!black,
  colorlinks = true
}

\newcommand{\geant}{\textsc{Geant4}\xspace}
\newcommand{\DaMaSCUS}{\textsc{DaMaSCUS}\xspace}
\newcommand{\Verne}{\textsc{Verne}\xspace}

\begin{document}

\title{Search for Daily Modulation of MeV Dark Matter Signals with DAMIC-M}



\author{I.\,Arnquist}
\affiliation{Pacific Northwest National Laboratory (PNNL), Richland, WA, United States} 

\author{N.\,Avalos}
\affiliation{Centro At\'{o}mico Bariloche and Instituto Balseiro, Comisi\'{o}n Nacional de Energ\'{i}a At\'{o}mica (CNEA), Consejo Nacional de Investigaciones Cient\'{i}ficas y T\'{e}cnicas (CONICET), Universidad Nacional de Cuyo (UNCUYO), San Carlos de Bariloche, Argentina}

\author{D.\,Baxter}\email[Now at Fermi National Accelerator Laboratory, Batavia, IL, United States]{}
\affiliation{Kavli Institute for Cosmological Physics and The Enrico Fermi Institute, The University of Chicago, Chicago, IL, United States}

\author{X.\,Bertou}
\affiliation{Centro At\'{o}mico Bariloche and Instituto Balseiro, Comisi\'{o}n Nacional de Energ\'{i}a At\'{o}mica (CNEA), Consejo Nacional de Investigaciones Cient\'{i}ficas y T\'{e}cnicas (CONICET), Universidad Nacional de Cuyo (UNCUYO), San Carlos de Bariloche, Argentina}

\author{N.\,Castell\'{o}-Mor}
\affiliation{Instituto de F\'{i}sica de Cantabria (IFCA), CSIC - Universidad de Cantabria, Santander, Spain}

\author{A.E.\,Chavarria}
\affiliation{Center for Experimental Nuclear Physics and Astrophysics, University of Washington, Seattle, WA, United States}

\author{J.\,Cuevas-Zepeda}
\affiliation{Kavli Institute for Cosmological Physics and The Enrico Fermi Institute, The University of Chicago, Chicago, IL, United States}

\author{A.\,Dastgheibi-Fard}
\affiliation{LPSC LSM, CNRS/IN2P3, Universit\'{e} Grenoble-Alpes, Grenoble, France}

\author{C.\,De Dominicis}
\affiliation{Laboratoire de physique nucl\'{e}aire et des hautes \'{e}nergies (LPNHE), Sorbonne Universit\'{e}, Universit\'{e} Paris Cit\'{e}, CNRS/IN2P3, Paris, France}

\author{O.\,Deligny}
\affiliation{CNRS/IN2P3, IJCLab, Universit\'{e} Paris-Saclay, Orsay, France}

\author{J.\,Duarte-Campderros}
\affiliation{Instituto de F\'{i}sica de Cantabria (IFCA), CSIC - Universidad de Cantabria, Santander, Spain}

\author{E.\,Estrada}
\affiliation{Centro At\'{o}mico Bariloche and Instituto Balseiro, Comisi\'{o}n Nacional de Energ\'{i}a At\'{o}mica (CNEA), Consejo Nacional de Investigaciones Cient\'{i}ficas y T\'{e}cnicas (CONICET), Universidad Nacional de Cuyo (UNCUYO), San Carlos de Bariloche, Argentina}

\author{N.\,Gadola}
\affiliation{Universit\"{a}t Z\"{u}rich Physik Institut, Z\"{u}rich, Switzerland}

\author{R.\,Ga\"{i}or}
\affiliation{Laboratoire de physique nucl\'{e}aire et des hautes \'{e}nergies (LPNHE), Sorbonne Universit\'{e}, Universit\'{e} Paris Cit\'{e}, CNRS/IN2P3, Paris, France}

\author{T.\,Hossbach}
\affiliation{Pacific Northwest National Laboratory (PNNL), Richland, WA, United States} 

\author{L.\,Iddir}
\affiliation{Laboratoire de physique nucl\'{e}aire et des hautes \'{e}nergies (LPNHE), Sorbonne Universit\'{e}, Universit\'{e} Paris Cit\'{e}, CNRS/IN2P3, Paris, France}

\author{B.~J.~Kavanagh}
\affiliation{Instituto de F\'{i}sica de Cantabria (IFCA), CSIC - Universidad de Cantabria, Santander, Spain}

\author{B.\,Kilminster}
\affiliation{Universit\"{a}t Z\"{u}rich Physik Institut, Z\"{u}rich, Switzerland}

\author{A.\,Lantero-Barreda}
\affiliation{Instituto de F\'{i}sica de Cantabria (IFCA), CSIC - Universidad de Cantabria, Santander, Spain}

\author{I.\,Lawson}
\affiliation{SNOLAB, Lively, ON, Canada }

\author{S.\,Lee}
\affiliation{Universit\"{a}t Z\"{u}rich Physik Institut, Z\"{u}rich, Switzerland}

\author{A.\,Letessier-Selvon}
\affiliation{Laboratoire de physique nucl\'{e}aire et des hautes \'{e}nergies (LPNHE), Sorbonne Universit\'{e}, Universit\'{e} Paris Cit\'{e}, CNRS/IN2P3, Paris, France}

\author{P.\,Loaiza}
\affiliation{CNRS/IN2P3, IJCLab, Universit\'{e} Paris-Saclay, Orsay, France}

\author{A.\,Lopez-Virto}
\affiliation{Instituto de F\'{i}sica de Cantabria (IFCA), CSIC - Universidad de Cantabria, Santander, Spain}

\author{K.~J.\,McGuire}
\affiliation{Center for Experimental Nuclear Physics and Astrophysics, University of Washington, Seattle, WA, United States}

\author{P.\,Mitra}
\affiliation{Center for Experimental Nuclear Physics and Astrophysics, University of Washington, Seattle, WA, United States}

\author{S.\,Munagavalasa}
\affiliation{Kavli Institute for Cosmological Physics and The Enrico Fermi Institute, The University of Chicago, Chicago, IL, United States}

\author{D.\,Norcini}
\affiliation{Kavli Institute for Cosmological Physics and The Enrico Fermi Institute, The University of Chicago, Chicago, IL, United States}
\affiliation{Department of Physics and Astronomy, Johns Hopkins University, Baltimore, MD,
United States}

\author{S.\,Paul}
\affiliation{Kavli Institute for Cosmological Physics and The Enrico Fermi Institute, The University of Chicago, Chicago, IL, United States}

\author{A.\,Piers}
\affiliation{Center for Experimental Nuclear Physics and Astrophysics, University of Washington, Seattle, WA, United States}

\author{P.\,Privitera}
\affiliation{Kavli Institute for Cosmological Physics and The Enrico Fermi Institute, The University of Chicago, Chicago, IL, United States}

\affiliation{Laboratoire de physique nucl\'{e}aire et des hautes \'{e}nergies (LPNHE), Sorbonne Universit\'{e}, Universit\'{e} Paris Cit\'{e}, CNRS/IN2P3, Paris, France}

\author{P.\,Robmann}
\affiliation{Universit\"{a}t Z\"{u}rich Physik Institut, Z\"{u}rich, Switzerland}

\author{S.\,Scorza}
\affiliation{LPSC LSM, CNRS/IN2P3, Universit\'{e} Grenoble-Alpes, Grenoble, France}

\author{M.\,Settimo}
\affiliation{SUBATECH, Nantes Universit\'{e}, IMT Atlantique, CNRS/IN2P3, Nantes, France}

\author{R.\,Smida}
\affiliation{Kavli Institute for Cosmological Physics and The Enrico Fermi Institute, The University of Chicago, Chicago, IL, United States}

\author{M.\,Traina}
\affiliation{Center for Experimental Nuclear Physics and Astrophysics, University of Washington, Seattle, WA, United States}
\affiliation{Laboratoire de physique nucl\'{e}aire et des hautes \'{e}nergies (LPNHE), Sorbonne Universit\'{e}, Universit\'{e} Paris Cit\'{e}, CNRS/IN2P3, Paris, France}

\author{R.\,Vilar}
\affiliation{Instituto de F\'{i}sica de Cantabria (IFCA), CSIC - Universidad de Cantabria, Santander, Spain}

\author{G.\,Warot}
\affiliation{LPSC LSM, CNRS/IN2P3, Universit\'{e} Grenoble-Alpes, Grenoble, France}

\author{R.\,Yajur}
\affiliation{Kavli Institute for Cosmological Physics and The Enrico Fermi Institute, The University of Chicago, Chicago, IL, United States}

\author{J-P.\,Zopounidis}
\affiliation{Laboratoire de physique nucl\'{e}aire et des hautes \'{e}nergies (LPNHE), Sorbonne Universit\'{e}, Universit\'{e} Paris Cit\'{e}, CNRS/IN2P3, Paris, France}

\collaboration{DAMIC-M Collaboration}

\date{\today}%

\begin{abstract}
Dark Matter (DM) particles with sufficiently large cross sections may scatter as they travel through Earth’s bulk. 
The corresponding changes in the DM flux give rise to a characteristic daily modulation signal in detectors sensitive to DM-electron interactions.
Here, we report results obtained from the first underground operation of the DAMIC-M prototype detector searching for such a signal from DM with MeV-scale mass. A model-independent analysis finds no modulation in the rate of 1$e^-$ events with sidereal period, where a DM signal would appear. 
We then use these data to place exclusion limits on DM in the mass range [0.53, 2.7] MeV/c$^2$ interacting with electrons via a dark photon mediator.
Taking advantage of the time-dependent signal we improve by $\sim$2 orders of magnitude on our previous limit obtained from the total rate of 1$e^-$ events, using the same data set. This daily modulation search represents the current strongest limit on DM-electron scattering via ultralight mediators for DM masses around 1 MeV/c$^2$. 

\end{abstract}


\maketitle

\textit{Introduction.---}
There are numerous astrophysical and cosmological observations supporting the existence of a non-baryonic and non-luminous form of matter~\cite{Bertone:2004pz}, known as dark matter (DM). 
Despite a wealth of experimental efforts to detect DM through its non-gravitational interactions~\cite{Gaskins:2016cha,Buchmueller:2017qhf,Billard:2021uyg}, its identity remains unknown.
DM particles with MeV-scale masses are viable candidates that appear in a range of natural, well-motivated scenarios~\cite{Boehm:2003ha,Hooper:2008im,Pospelov:2007mp,Hooper:2008im,Chu:2011be,Knapen:2017xzo}.
However, their discovery remains challenging since such light particles do not have sufficient kinetic energy to produce detectable nuclear recoils in direct-detection experiments~\cite{Essig:2011nj}. Instead, DM-\textit{electron} scattering would produce eV-scale electronic recoils, which would be observed in semiconductor-based detectors with $\mathcal{O}$(1 eV) ionization thresholds~\cite{Essig:2015cda}. 
To date, the sensitivity of semiconductor searches~\cite{SuperCDMS:2018mne,DAMIC:2019dcn,SENSEI:2020dpa,EDELWEISS:2020fxc,DAMIC-M:2023gxo} has been partly limited by background levels~\cite{Fuss:2022fxe}. A time-dependent DM signal detected above a time-independent background can significantly increase the sensitivity of these searches.

Various phenomena give rise to a daily modulation in the flux of DM,
including the variation of the velocity of the lab frame due to the Earth's rotation~\cite{Kouvaris:2015xga} and gravitational focusing of the DM flux by the Earth~\cite{Sikivie:2002bj,Alenazi:2006wu,Kouvaris:2015xga}.
Furthermore, if the DM scattering cross-section is sufficiently large, the velocity distribution of DM particles at the detector may be distorted by their interactions in the Earth~\cite{Collar:1992qc,Collar:1993ss,Hasenbalg:1997hs}. Over the course of a sidereal day, the position of the detector rotates with respect to the incoming DM flux. Thus, the DM particles will travel a greater or smaller distance across the Earth at different times, leading to a daily modulation. 

Most experimental searches for daily modulations~\cite{damadaily,luxdaily,pandadaily,prospectdaily} have targeted DM particles interacting with nuclei.
Here, we focus on models with MeV-scale DM, which may be coupled to Standard Model particles via a kinetically-mixed dark photon~\cite{Holdom1986,1984PhLB..136..279G}, and thus interact with electrons. 
For DM-electron scattering cross-sections currently within reach of semiconductor searches, interactions in the  Earth may be substantial~\cite{Emken:2019tni}, which motivate the search for a daily modulated signal on top of a constant background~\cite{Kouvaris:2014lpa,Avalos_2021}. 

In this letter, we present a search for a daily modulated signal with a prototype of the DAMIC-M (Dark Matter in CCDs at Modane) experiment~\cite{damic_idm2022}.  
We first perform a model-independent search for modulations over a wide range of periods. 
We then report limits on DM particles with MeV-scale masses interacting with electrons, which incorporate, for the first time with an underground detector, a time-dependent analysis to provide a powerful discriminant between signal and background.

\textit{Setup and data.---}The DAMIC-M experiment will feature $\sim$200 skipper charge-coupled devices (CCDs), for a target silicon mass of $\sim$700 g, installed under the French Alps at the Laboratoire Souterrain de Modane (LSM). DAMIC-M is sensitive to a single-ionization charge signal thanks to the extremely low level of dark current and the sub-electron readout noise of the devices. 
By operating prototype CCDs at LSM in the Low Background Chamber (LBC) we obtained world-leading limits on sub-GeV DM particles interacting with electrons~\cite{DAMIC-M:2023gxo}. The daily modulation results reported here are obtained with the same apparatus and data set, which are summarized in the following. 

Two high-resistivity, ($>$10\,k$\Omega$\,cm) $n$-type silicon CCDs~\cite{Holland:2002zz,Holland:2003zz,Holland:2009zz} are installed in the LBC. Each CCD has $6144\times 4128$ pixels (pixel area $15\times 15$ $\mu$m$^2$) and is $670$ $\mu$m thick, for a mass of $\sim$$9$ g. The detectors are mounted in a high-purity copper box, which also shields infrared radiation. A lead shield of $\sim$$7.5$ cm thickness, with the innermost 2 cm of ancient origin~\cite{ALESSANDRELLO1991106}, encloses the copper box inside the LBC cryostat. An additional 15 cm of lead and 20 cm of high density polyethylene surround the cryostat to shield from environmental radiation. The measured background between 1 and 20~keV is $\sim$$10$ events/keV/kg/day, consistent with expectations from a \geant\,\cite{geant4} Monte Carlo simulation of the apparatus.
 The CCDs are kept at a temperature of $\sim130$ K inside the vacuum cryostat at $\sim$$5\times 10^{-6}$ mbar. 

Charge carriers produced by ionizing radiation in the silicon bulk of the CCD drift towards the $x$-$y$ plane where pixels are located. Thermal diffusion while drifting results in a spatial variance, $\mathrm{\sigma^{2}_{xy}}$, of the charge collected in contiguous pixels~\cite{PhysRevD.94.082006}. In the readout process, the pixel charge is moved across the array by appropriate voltage clocking and read out serially by two amplifiers (referred as U and L) located at the corners of the CCD at the end of the serial register. These skipper amplifiers~\cite{skipper,Chandler1990zz,Tiffenberg:2017aac} allow for multiple non-destructive measurements (NDCMs) of the pixel charge.  Sub-electron resolution is achieved by averaging a sufficient number ($N_\mathrm{skips}$) of NDCMs, with the resolution improving as $1/\sqrt{N_\mathrm{skips}}$. The performance of DAMIC-M CCDs is detailed in Ref.\,\cite{Compton-DAMICM}.

For this analysis we use 63 days of uninterrupted data-taking beginning on June 8th 2022 (the SR2 data set in Ref.~\cite{DAMIC-M:2023gxo}). Relevant CCD operating parameters include $N_\mathrm{skips}=650$, corresponding to a pixel charge resolution of $\sim$$0.2\,e^-$, and a $\mathrm{10\times10}$ pixel binning.
In subsequent text, the term pixel is used to describe a $10\times10$ bin of pixels from which charge is summed before being read out. To minimize dark current counts, only a fraction of the CCD is read out ($640\times110$ pixels), and its charge is cleared between consecutive images.

Image data reduction and pixel selection criteria are detailed in Ref.~\cite{DAMIC-M:2023gxo}. In brief, 
images are first calibrated from the fitted position of the 0$e^-$ and 1$e^-$ peaks in the pixel charge distribution.
Then contiguous pixels with charge are joined into clusters. Those with total charge $>$7$e^-$ are excluded from further analysis since they are unlikely to be produced by sub-GeV DM interactions with electrons. For each cluster we also exclude 10 trailing pixels in the horizontal and vertical directions to account for charge transfer inefficiencies. Defects in the CCDs may release charge during the readout, appearing in the images as ``hot" pixels and columns~\cite{2001sccd.book.....J}. These are identified by their increased rate of pixels with 1$e^{-}$, and rejected. Lastly, we exclude one of the prototype CCDs due to the presence of several charge traps in the serial register (identified by a decreased rate of pixels with 1$e^{-}$). A portion of the remaining CCD is also excluded for the same reason. 
The final integrated exposure after applying these selection criteria is 39.97 g-days, distributed over $N_\mathrm{im} = 8779$ images.

For each image $i$ the charge distribution of the selected pixels is fitted with the sum of two Gaussian functions corresponding to the 0$e^-$ and 1$e^-$ peaks. The number of pixels with 1$e^-$ charge, $N_1^i$, and its uncertainty $\sigma_{N_1}^i$ are then obtained from the fitted parameters. The U and L sides of the image are fitted independently to account for small differences in the calibration and resolution of the amplifiers. We find the value of $N_1^i$ to be $\sim$80 and $\sim$30 pixels/image for the U and L amplifier, respectively. The difference in counts is due to a large portion of the L side being excluded by the serial register trap criteria. Corresponding rates of 1$e^-$, $R_1^i$, and associated uncertainties $\sigma_{R_1}^i$ are then obtained dividing $N_1^i$ and $\sigma_{N_1}^i$ by the effective image exposure time ($t_\mathrm{exp}=0.00356$ days) and target mass after pixel selection.
For each image the relative uncertainty $\sigma_{R_1}^i/R_1^i$ is $\sim$$13 (21)\%$ for the U (L) side. This set of 1$e^-$ rates, measured every $\sim$10 min for 63 consecutive days, is used to search for a daily modulation.

\begin{figure}
     \centering
     \includegraphics[width=0.48\textwidth]{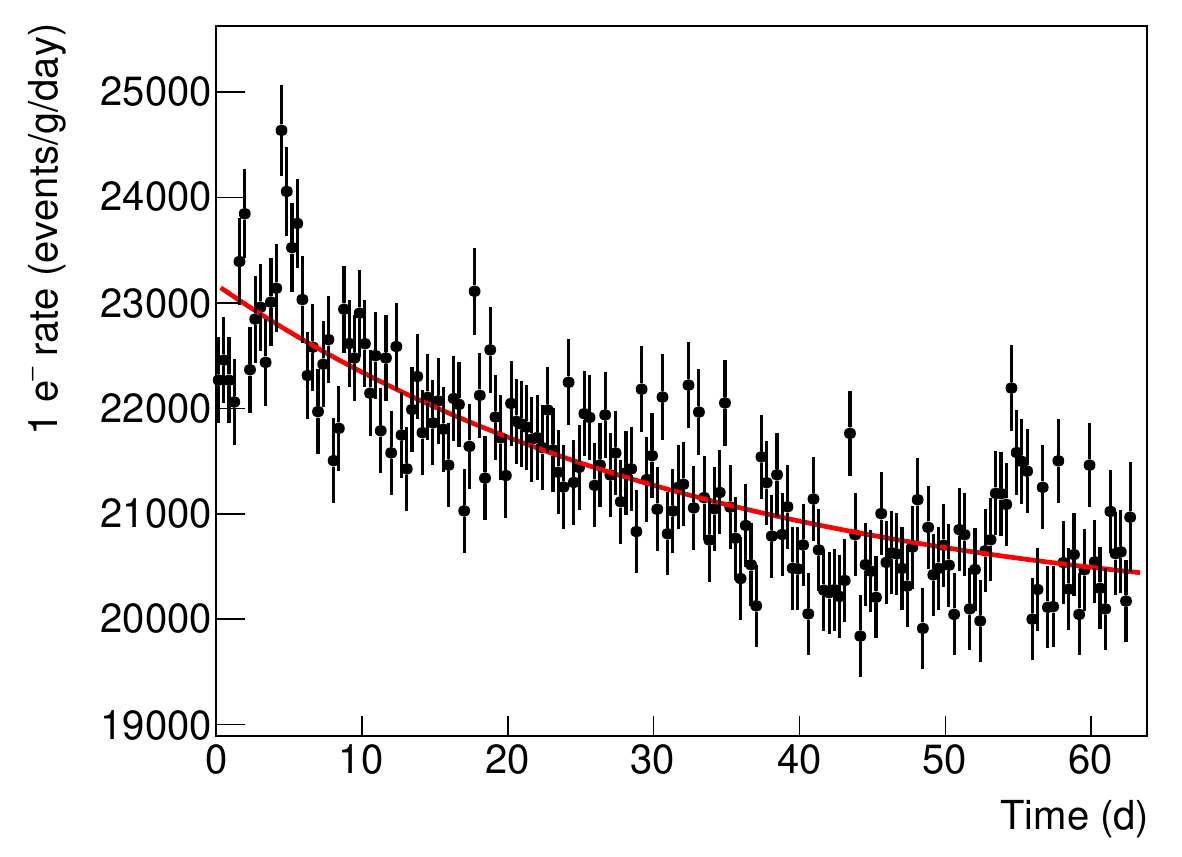}
     \caption{Measured rate of 1$e^-$, $R_1^i$, as a function of time for the U-side. Each data point is the rate averaged over 50 images. 
     The red line is the result of the fit of a constant plus an exponential. 
     }
     \label{fig:rate_time}
\end{figure}

The rate $R_1^i$ for the U side is shown as a function of time in Fig.~\ref{fig:rate_time}.  The rate is well-parameterized by a constant plus an exponential with time decay constant $\tau\sim33.5$ days. The L side has the same behavior. 
Since $\tau \gg$ 1~day, a daily modulation analysis is largely unaffected. The observed time dependence is consistent with the dark current stabilizing over time, a characteristic of these types of devices~\cite{2001sccd.book.....J}. 

\textit{Model-independent search for a daily modulation.---}We first perform a model-independent analysis by fitting the measured rates $R_1^i$ to a function $F(t)=Be^{-t/\tau} + C + A\cos((2\pi(t - \phi)/T)$ where $B,C~$ and $\tau$ parameterize the exponential time dependence observed in the data, and $A,~T$ and $\phi$ are the modulation amplitude, period and phase. This simplified model approximates a periodic signal with the first term of its Fourier expansion. We use a binned likelihood method to test the null hypothesis, $i.e.$ absence of modulation,  corresponding to $A=0$. The likelihood function is
\begin{equation}\label{eq:binned_likelihood}
 \mathcal{L}(\theta)=\prod_{i=1}^{N_\mathrm{im}}\frac{1}{\sqrt{2\pi}\sigma_{R_1}^{i}}\exp\left\{-\frac{1}{2}\left(\frac{R_1^i-F(t_i|\theta)}{\sigma_{R_1}^i}\right)^2\right\}\,,
\end{equation}
where $\theta=\{A, B, C, \tau, T, \phi\}$ is the set of model parameters 
and $t_i$ is the time~\footnote{The time of each image was recorded in Coordinated Universal Time (UTC)} when image $i$ was taken. The spread in the arrival time of the events within each image has negligible impact for the range of periods explored. 
The likelihood of the no-modulation hypothesis $H_0$ is found by maximizing Eq.~\eqref{eq:binned_likelihood} under the constraint $A=0$, $\mathcal{L}_{H_0} = \sup_{A=0}\mathcal{L(\theta)}$, while the global maximum for a given period $T$ is $\mathcal{L}_{H_1} = \sup_{\theta}\mathcal{L}(\theta)$. The likelihood-ratio test statistic for the $H_0$ hypothesis is given by $t_{q} = -2\ln\left(\mathcal{L}_{H_0}/\mathcal{L}_{H_1}\right)$.
The local significance of a departure from the $H_0$ hypothesis for any given period $T$ can then be quantified by the value of $t_{q}$~\cite{Cowan:2010js}. 
\begin{figure}[t]
     \centering
     \includegraphics[width=0.48\textwidth]{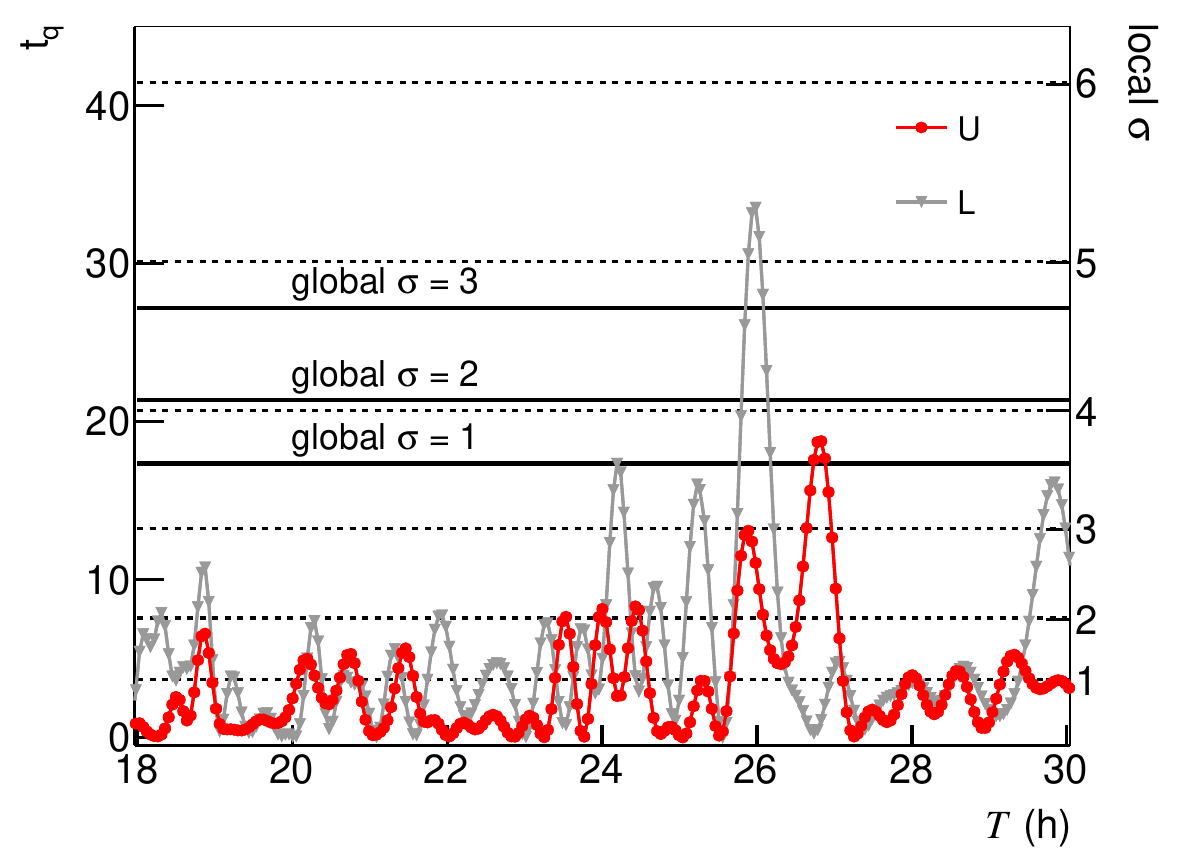} 
     \caption{
      Model-independent search for daily modulation: the test statistic $t_q$ and corresponding local significance as a function of the modulation period $T$ in a range around 24\,h; the global significance levels for 1, 2 and 3 $\sigma$ are also shown as horizontal lines. Both local and global significances are one-sided. See text for details.} 
      
     \label{fig:model_independent_plot}
\end{figure}
In Fig.~\ref{fig:model_independent_plot} we show the value of the test statistic and the corresponding local and global significance as a function of $T$ in a range around 24\,h. Global significances are calculated by examining the distribution of the maximum value of the test statistic under $H_0$ ~\cite{Gross:2010qma}, accounting for the search range T=1-48\,h. The measured rate is consistent with the null hypothesis (no-modulation) for the sidereal period (T=23.93\,h), where a DM signal would appear. The same conclusion is drawn for all other periods in the search range, with the only deviation found for T=26\,h in the L-side. This deviation has no effect on the DM-$e^{-}$ scattering search (see Supplemental Material for details). 

\textit{DM-$e^{-}$ scattering.---}
In this section, we introduce a new approach for constraining DM-$e^{-}$ scattering, which exploits the expected daily modulation due to DM interactions within the Earth. The rate of DM-$e^-$ scattering in a semiconductor is obtained by convolving the scattering cross-section with the DM velocity distribution at the detector $f(\mathbf{v}, t)$ and with the crystal form factor $f_\mathrm{c}$~\cite{Essig:2015cda,Lee:2015qva,Griffin:2021znd}: 
\begin{align}
\label{eq:dR_dEe}
\begin{split}
    \frac{\mathrm{d}R}{\mathrm{d}E_e} \propto \bar{\sigma}_e \int \frac{\mathrm{d}q}{q^2}\left[\int\frac{f(\mathbf{v}, t)}{\mathrm{v}}
    \,\mathrm{d^3v}\right]\left|F_\mathrm{DM}(q)\right|^2\left|f_\mathrm{c}(q,E_e)\right|^2\,.
\end{split}
\end{align}
Here, $\bar{\sigma}_e$ is a reference DM-$e^-$ scattering cross-section, $E_e$ is the energy deposited as ionization and $q$ is the momentum transfer. We consider a model of DM which couples to Standard Model particles via a kinetically mixed dark photon. In this case, the DM form factor $F_\mathrm{DM} = (\alpha m_e/q)^n$ depends on the mass of the mediator $m_{A^\prime}$, with $n=0$ for a heavy mediator ($m_{A^\prime} \gg \alpha \, m_e$) and $n=2$ for ultralight mediators ($m_{A^\prime} \ll \alpha \, m_e$)~\cite{Emken:2019tni}.

The velocity distribution of DM is expected to follow a Maxwell-Boltzmann distribution in the Galactic frame~\cite{Green:2017odb}\footnote{We use the Standard Halo Model~\cite{PhystatDM} parameters with the Earth's velocity fixed to its average value during the data-taking period, $v_{E}=263$~km/s.}. 
The velocity distribution at the detector is obtained by boosting into the rest-frame of the laboratory, which is moving at a velocity $\mathbf{v}_\mathrm{lab}(t) = \mathbf{v}_\odot + \mathbf{v}_\mathrm{E}(t) + \mathbf{v}_\mathrm{rot}(t)$. Here, $\mathbf{v}_\odot$ is the velocity of the Sun in the Galactic frame, $\mathbf{v}_\mathrm{E}(t)$ the Earth's velocity in the Solar frame~\cite{McCabe:2013kea}, and $\mathbf{v}_\mathrm{rot}(t)$ is the rotational velocity of the lab in an Earth-centered inertial frame~\cite[App.~A]{Emken:2017qmp}. The velocity distribution $f(\mathbf{v}, t)$ at the detector is therefore always time-dependent due to the Earth's motion.

If the DM-scattering cross section is sufficiently large, however, the flux of DM may also be distorted by scattering of particles in the Earth before they reach the detector. The velocity distribution then depends on the isodetection angle $\Theta$~\cite{Emken:2017qmp}, defined as the angle between the local zenith and the direction of the mean DM flux $\langle \mathbf{v}_\chi \rangle = - \mathbf{v}_\mathrm{lab}$. For $\Theta = 0^\circ$, the mean DM flux comes from directly above the detector, while at $\Theta = 180^\circ$, it comes from directly below. 
The isodetection angle oscillates over a sidereal day with the Earth's rotation, and thus the expected DM flux.
During the data-taking period the isodetection angle at LSM ($45.2^\circ$ N) varied in the range $\Theta \in [0^\circ, 92^\circ]$. 

Depending on the regime of interest, a number of formalisms have been developed to estimate the DM flux at the detector as a function of $\Theta$, taking into account Earth-scattering (e.g.~\cite{Starkman:212913,Kouvaris:2014lpa,Kavanagh:2016pyr,Mahdawi:2017cxz,Emken:2017qmp,Hooper:2018bfw,Mahdawi:2018euy}).
Here, we employ a modified version of the \Verne code~\cite{Kavanagh:2017cru,Verne}\footnote{Note that \Verne uses an alternative convention for the isodetection angle $\gamma = 180^\circ - \Theta$.}, which assumes that light DM particles travel along straight-line trajectories and either continue unaffected or are reflected back along their incoming path when they scatter~\cite{Lantero_inprep}. 
DM scattering with Earth nuclei, implemented in \Verne with charge-screening~\cite{Emken:2019tni},  is the dominant process for daily modulation effects.
Since each $t_i$ corresponds to a $\Theta(t_i)$~\footnote{Note that the isodetection angle corresponding to a given time in a sidereal day changes by several degrees during the data taking period}, this code allows us to efficiently calculate $f(\mathbf{v}, t_i)$ over a wide range of parameter space.

We search for a daily modulation of the rate $R_1^i$ by performing a likelihood fit with Eq.~\eqref{eq:binned_likelihood}, where $F$ includes both the background and signal model. 
For the background we use a Poisson model, $\textup{Pois}(n_q|\lambda)$, which gives the probability of measuring $n_q$ charges in a pixel given a dark current $\lambda$ (in units of counts/image).
As shown in Fig.~\ref{fig:rate_time}, the dark current in our data stabilizes with an exponential decrease in time.  Thus, we model the time-dependent dark current as $\lambda(t_i)=\lambda_A\cdot \exp(-t_{i}/\tau) + \lambda_\mathrm{eq}$, where $\lambda_A$ and $\tau$ are the amplitude and time constant of the exponential decay of the dark current, and $\lambda_\mathrm{eq}$ is its value once stabilized. 
The signal model, $S(n_q|m_\chi,\bar{\sigma}_e,t_{i})$, gives the probability that $n_q$ charges are measured in a pixel for DM particles of mass $m_\chi$ and cross section $\bar{\sigma}_e$; the time dependence in $S$ accounts for the expected daily modulation. The fitting function in Eq.~\eqref{eq:binned_likelihood} is then: 
\begin{equation}\label{eq:N_electrons_im}
  F(t_i|\theta)   = \frac{1}{t_\mathrm{exp} m_\mathrm{pix}} \sum_{j=0}^{1} \textup{Pois}(1-j|\lambda(t_{i})) S(j|m_\chi,\bar{\sigma}_e,t_{i})\,,
\end{equation}
where $m_\mathrm{pix}$ is the mass of one CCD pixel and $\theta = \{\lambda_A,\tau,\lambda_\mathrm{eq},\bar{\sigma}_e\}$ are free parameters in the fit. 
\begin{figure}
     \centering
     \includegraphics[width=0.48\textwidth]{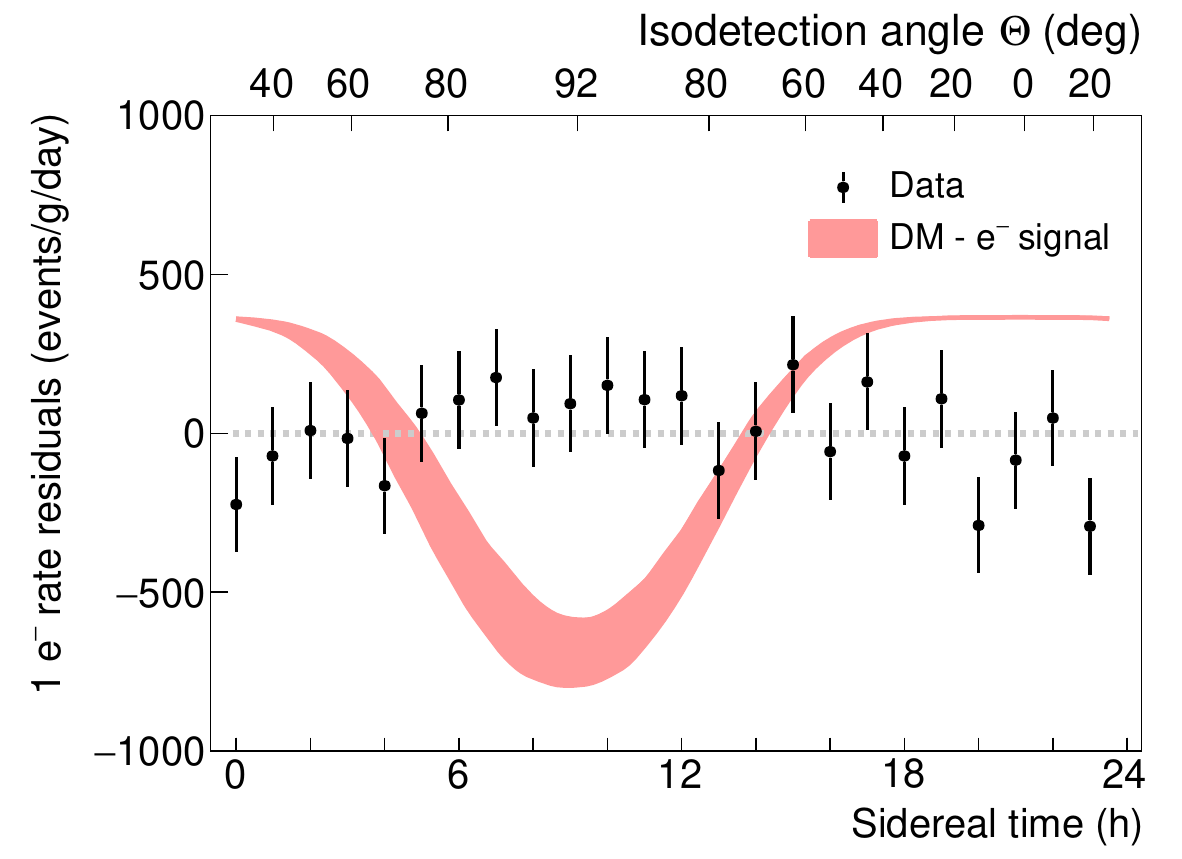}

     \caption{ U-side residual rate after the subtraction of the best-fit background-only model, binned as a function of local apparent sidereal time. 
     As a reference, the upper $x$-axis gives the isodetection angle $\Theta(t)$ for the first day of data taking.
     Each data point is obtained from the average of $\sim$\,300 images.
     The light red band shows the expected signal (minus its time-average) for a DM particle of mass 1\,MeV/c$^2$, $\bar{\sigma}_e=10^{-32}$\,cm$^2$ interacting via an ultralight dark photon mediator. The signal is shown as a band because a given value of the sidereal time may correspond to a range of different $\Theta$ values, depending on the varying Earth velocity over the year. This gives rise to a time shift and amplitude change of the signal during the data taking period.
     }
     \label{fig:isoangle_time}
\end{figure} 
\begin{figure*}
    \centering
	\hspace{-0.2cm}
        \includegraphics[width=0.48\textwidth]{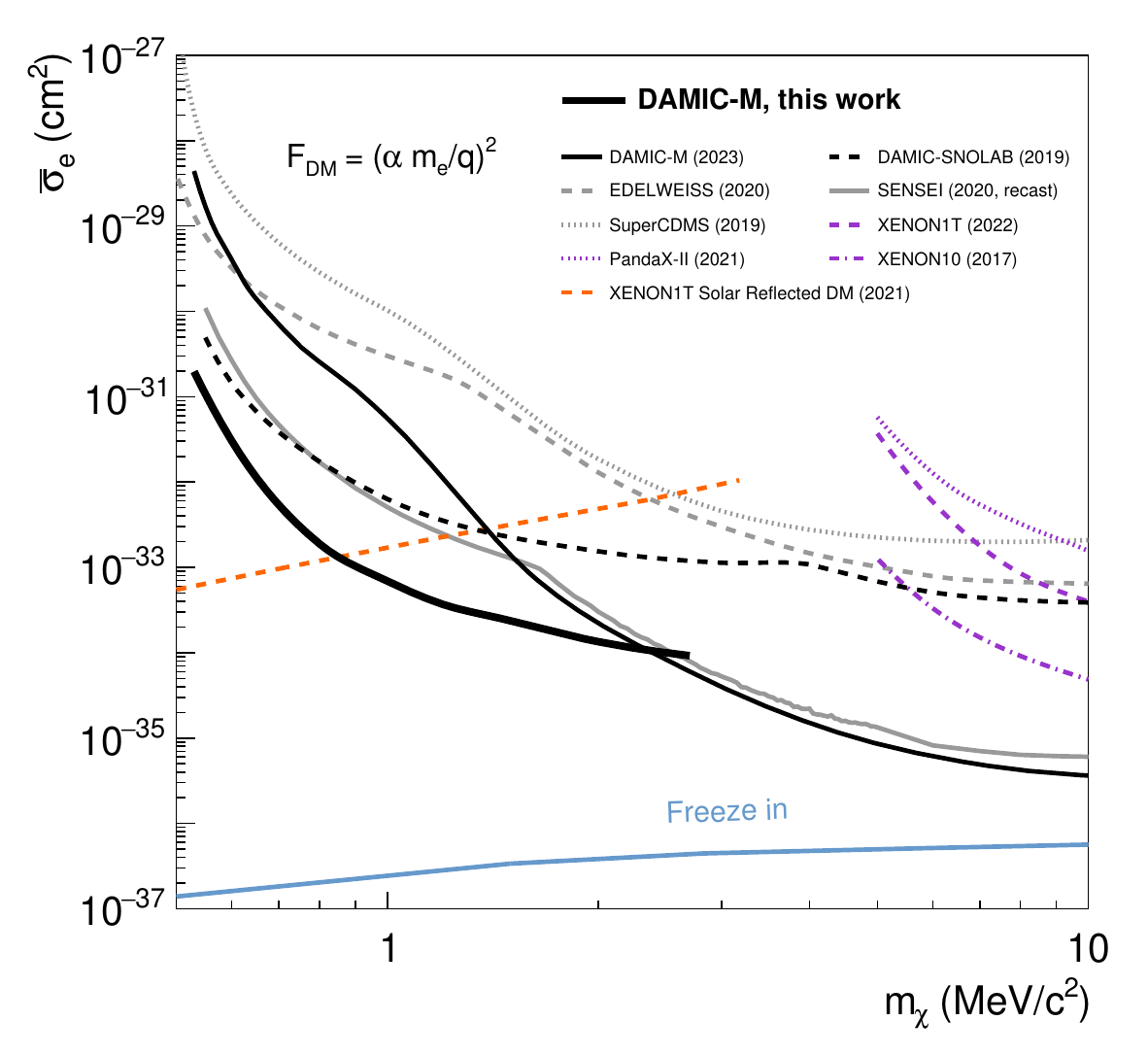}
        \hspace{0.6cm}
   	\includegraphics[width=0.48\textwidth]{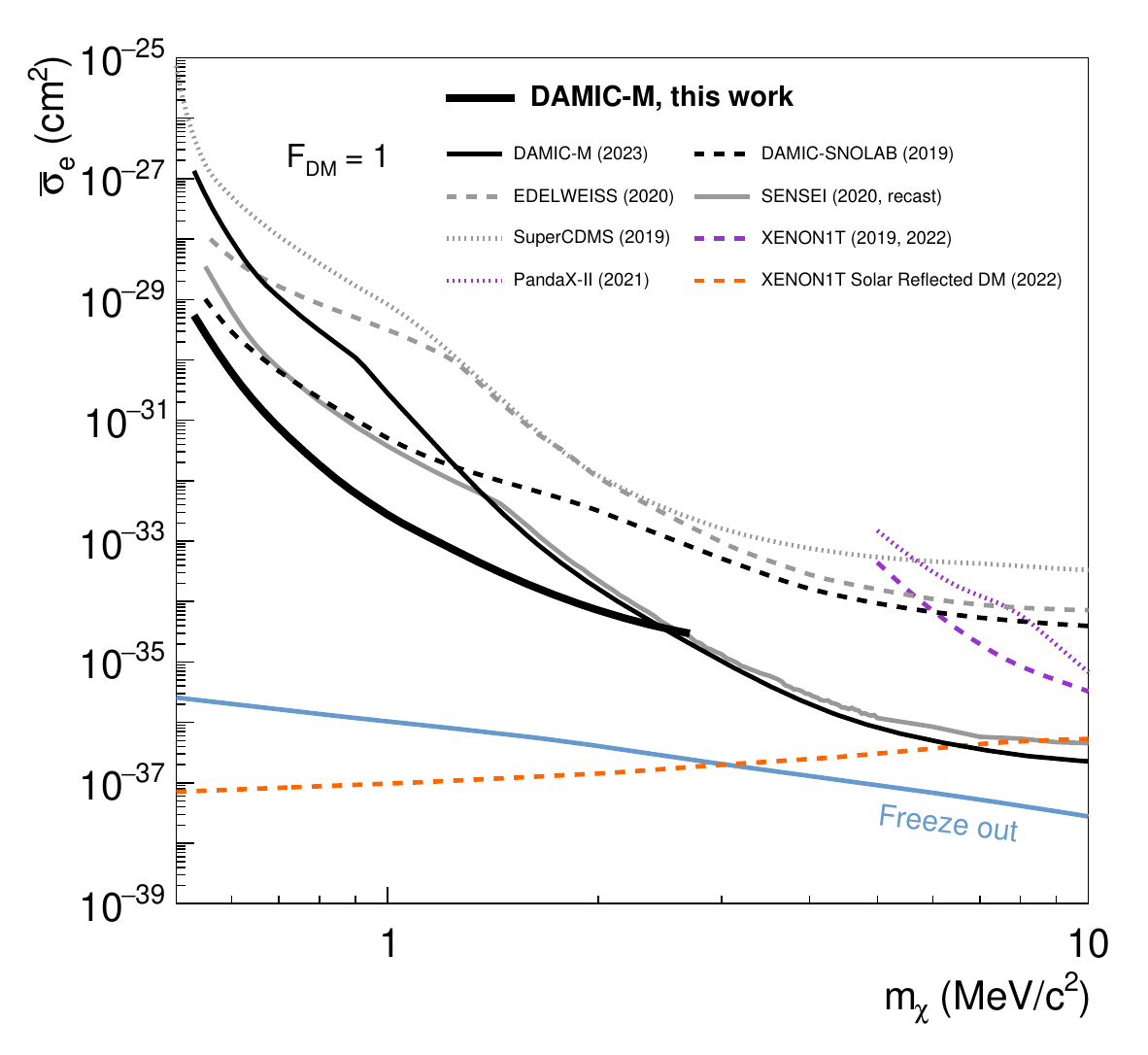}
    \caption{DAMIC-M 90\% C.L. upper limits (solid thick black) on DM-electron interactions through an ultralight (left) and heavy (right) dark photon mediator obtained from the daily modulation analysis. Also shown are previous limits from DAMIC-M~\cite{DAMIC-M:2023gxo} (solid black) and other experiments: DAMIC-SNOLAB~\cite{DAMIC:2019dcn} (dashed black); SENSEI~\cite{SENSEI:2020dpa,SENSEIrecast} (solid gray); EDELWEISS~\cite{EDELWEISS:2020fxc} (dashed gray); SuperCDMS~\cite{SuperCDMS:2018mne} (dotted gray); 
    XENON1T combined result from~\cite{xenon1t2019,xenon1t-SE-2022} (dashed violet); PandaX-II~\cite{PandaX-2021} (dotted violet); a limit obtained from XENON10 data in Ref.~\cite{xenon10_2017} (dash-dotted violet); and a limit obtained from XENON1T data considering ``solar reflected DM" (dashed orange) from Ref.~\cite{SolarReflectedDM} (left) and  Ref.~\cite{PhysRevD.105.063020} (right). Theoretical expectations assuming a DM relic abundance from freeze-in and freeze-out mechanisms are also shown in light blue~\cite{USCosmicVisions2017}. 
    }
\label{fig:upperlimits}
 \end{figure*}
The signal model $S$ is derived with the following procedure. First, the rate as a function of the energy $E_e$ deposited in the DM interaction is computed with Eq.~\eqref{eq:dR_dEe} using $\texttt{QEDark}$~\cite{Essig:2015cda} for the crystal form factor $f_\mathrm{c}$, which encodes the electronic response of the target. Then, $E_e$ is converted to electron-hole pairs with a semi-empirical model based on probabilities $P_\mathrm{pair}(n_q|E_e)$ from the charge yield model in Ref.~\cite{PhysRevD.102.063026}. To include the detector response, we perform a Monte Carlo simulation of the charge diffusion process, with $\sigma_{xy}^2$ obtained from a surface laboratory calibration with cosmic rays~\cite{PhysRevD.105.062003,DAMIC-M:2023gxo}. Lastly, a $10\times10$ binning of the simulated pixel array is applied to match the data-taking conditions.

We minimize the joint log-likelihood for the U and L sides at fixed DM masses between 0.53 and 2.7 MeV/c$^2$.
To illustrate the sensitivity of the method, we show in Fig.~\ref{fig:isoangle_time} the residual rate after background subtraction, obtained by fitting the U-side data to the background model only.
Also shown is the expectation for a DM particle of mass 1\,MeV/c$^2$ interacting via an ultralight mediator with $\bar{\sigma}_e=10^{-32}$\,cm$^2$ (close to the current best limit on the cross section at this mass).

The fit finds no preference for a DM signal at any mass. 
Consequently, we derive exclusion limits using the approach of Ref.\,\cite{asymptoticlikelihood} and the profile likelihood ratio test statistic, $t_\sigma = -2\log\lambda(\sigma)$ where $\lambda(\sigma)$ is the profile likelihood ratio, at each DM mass.
The corresponding 90\% C.L. exclusion limits for ultralight (left) and heavy (right) mediators are shown in Fig.~\ref{fig:upperlimits}. These limits fall within the expected 95.4\% sensitivity band as estimated by Monte Carlo simulations. Also shown are previous results from DAMIC-M and other direct detection experiments.
Notably, the daily modulation analysis improves up to $\sim$2 orders of magnitude the previous DAMIC-M limits~\cite{DAMIC-M:2023gxo} obtained with the same data set. Note that the excluded region does not extend to indefinitely large cross sections, as Earth-scattering effects eventually lead to a complete attenuation of the DM flux. For the light mediator, this upper bound occurs around $\sigma_\mathrm{e} \approx 10^{-27}\,\mathrm{cm}^2$ over the mass range of interest. For the heavy mediator, this upper bound is at $\sigma_\mathrm{e} \approx 10^{-27}\,\mathrm{cm}^2$ near $m_\chi = 1\,\mathrm{MeV}/c^2$, decreasing to $\sigma_\mathrm{e} \approx 10^{-30}\,\mathrm{cm}^2$ at $m_\chi = 10\,\mathrm{MeV}/c^2$~\cite{Emken:2019tni}. Cross-checks of the analysis are included in the Supplemental Material.

For DM masses $\le 2.7$\,MeV/c$^2$ the signal is overwhelmingly comprised of 1$e^-$. At higher masses the 2$e^-$ signal becomes relevant in establishing the exclusion limit. However, the existing cross section constraints result in a $R_2^i$ daily modulation amplitude that is too small to be detectable with the statistics of the current data set. 
For this reason, we do not perform a full daily modulation analysis of $R_2^i$. 

\textit{Conclusions.---}
This DAMIC-M search for DM particles interacting with electrons excludes unexplored regions of parameter space between 0.53 and 2.7 MeV/c$^2$. 
This is the first time that the daily modulation due to scattering of DM particles in the Earth, before they reach the detector, constrains DM-electron interactions. When combined with our previous limits, DAMIC-M provides the current best constraints from searches for a non-relativistic flux of DM particles incident on Earth, over the mass ranges $[ 0.53, 1000 ]$ MeV/c$^2$ and $[ 0.53, 15.1 ]$ MeV/c$^2$ for ultralight and heavy mediator interactions, respectively \footnote{Complementary searches for semi-relativistic DM fluxes from Solar
reflection~\cite{SolarReflectedDM,PhysRevD.105.063020}, and cosmological~\cite{Krnjaic:2019dzc,Giovanetti:2021izc} and stellar evolution constraints~\cite{Davidson:2000hf,Chang:2018rso} also place limits in this mass range.}.
In addition, a model-independent search for modulations with period close to 24\,h demonstrates the stability of our detector, allowing for further improvements in time-dependent searches for a DM signal.

\textit{Acknowledgments.---}We thank the LSM for their support in the installation and operation of the detector underground.
The DAMIC-M project has received funding from the European Research Council (ERC) under the European Union's Horizon 2020 research and innovation programme Grant Agreement No. 788137, and from NSF through Grant No. NSF PHY-1812654. 
The work at University of Chicago and University of Washington was supported through Grant No. NSF PHY-2110585. This work was supported by the Kavli Institute for Cosmological Physics at the University of Chicago through an endowment from the Kavli Foundation. 
We thank the College of Arts and Sciences at University of Washington for 
contributing the first CCDs to the DAMIC-M project.
IFCA was supported by project PID2019-109829GB-I00 funded by MCIN/ AEI /10.13039/501100011033.
BJK acknowledges funding from the Ram\'on y Cajal Grant RYC2021-034757-I, financed by MCIN/AEI/10.13039/501100011033 and by the European Union ``NextGenerationEU"/PRTR.
The Centro At\'{o}mico Bariloche group is supported by ANPCyT grant PICT-2018-03069.
The University of Z\"{u}rich was supported by the Swiss National Science Foundation.
The CCD development work at Lawrence Berkeley National Laboratory MicroSystems Lab was supported in part by the Director, Office of Science, of the U.S. Department of Energy under Contract No. DE-AC02-05CH11231. We thank Teledyne DALSA Semiconductor for CCD fabrication. 
We acknowledge Santander Supercomputing support group at the University of Cantabria who provided access to the supercomputer Altamira at the Institute of Physics of Cantabria (IFCA-CSIC), member of the Spanish Supercomputing Network, for performing simulations.

\bibliographystyle{style/h-physrev}

\bibliography{apssamp}

\providecommand{\noopsort}[1]{}\providecommand{\singleletter}[1]{#1}%
\begin{thebibliography}{10}

\bibitem{Bertone:2004pz}
G.~Bertone, D.~Hooper, and J.~Silk,
\newblock Phys. Rept. {\bf 405}, 279 (2005), hep-ph/0404175.

\bibitem{Gaskins:2016cha}
J.~M. Gaskins,
\newblock Contemp. Phys. {\bf 57}, 496 (2016), 1604.00014.

\bibitem{Buchmueller:2017qhf}
O.~Buchmueller, C.~Doglioni, and L.~T. Wang,
\newblock Nature Phys. {\bf 13}, 217 (2017), 1912.12739.

\bibitem{Billard:2021uyg}
J.~Billard {\em et~al.},
\newblock Rept. Prog. Phys. {\bf 85}, 056201 (2022), 2104.07634.

\bibitem{Boehm:2003ha}
C.~Boehm, P.~Fayet, and J.~Silk,
\newblock Phys. Rev. D {\bf 69}, 101302 (2004), hep-ph/0311143.

\bibitem{Hooper:2008im}
D.~Hooper and K.~M. Zurek,
\newblock Phys. Rev. D {\bf 77}, 087302 (2008), 0801.3686.

\bibitem{Pospelov:2007mp}
M.~Pospelov, A.~Ritz, and M.~B. Voloshin,
\newblock Phys. Lett. B {\bf 662}, 53 (2008), 0711.4866.

\bibitem{Chu:2011be}
X.~Chu, T.~Hambye, and M.~H.~G. Tytgat,
\newblock JCAP {\bf 05}, 034 (2012), 1112.0493.

\bibitem{Knapen:2017xzo}
S.~Knapen, T.~Lin, and K.~M. Zurek,
\newblock Phys. Rev. D {\bf 96}, 115021 (2017), 1709.07882.

\bibitem{Essig:2011nj}
R.~Essig, J.~Mardon, and T.~Volansky,
\newblock Phys. Rev. D {\bf 85}, 076007 (2012), 1108.5383.

\bibitem{Essig:2015cda}
R.~Essig {\em et~al.},
\newblock JHEP {\bf 05}, 046 (2016), 1509.01598.

\bibitem{SuperCDMS:2018mne}
SuperCDMS Collaboration, R.~Agnese {\em et~al.},
\newblock Phys. Rev. Lett. {\bf 121}, 051301 (2018), 1804.10697,
\newblock [Erratum: Phys.Rev.Lett. 122, 069901 (2019)].

\bibitem{DAMIC:2019dcn}
DAMIC Collaboration, A.~Aguilar-Arevalo {\em et~al.},
\newblock Phys. Rev. Lett. {\bf 123}, 181802 (2019), 1907.12628.

\bibitem{SENSEI:2020dpa}
SENSEI Collaboration, L.~Barak {\em et~al.},
\newblock Phys. Rev. Lett. {\bf 125}, 171802 (2020), 2004.11378.

\bibitem{EDELWEISS:2020fxc}
EDELWEISS Collaboration, Q.~Arnaud {\em et~al.},
\newblock Phys. Rev. Lett. {\bf 125}, 141301 (2020), 2003.01046.

\bibitem{DAMIC-M:2023gxo}
DAMIC-M Collaboration, I.~Arnquist {\em et~al.},
\newblock Phys. Rev. Lett. {\bf 130}, 171003 (2023), 2302.02372.

\bibitem{Fuss:2022fxe}
P.~Adari {\em et~al.},
\newblock SciPost Phys. Proc. {\bf 9}, 001 (2022), 2202.05097.

\bibitem{Kouvaris:2015xga}
C.~Kouvaris and N.~G. Nielsen,
\newblock Phys. Rev. D {\bf 92}, 075016 (2015), 1505.02615.

\bibitem{Sikivie:2002bj}
P.~Sikivie and S.~Wick,
\newblock Phys. Rev. D {\bf 66}, 023504 (2002), astro-ph/0203448.

\bibitem{Alenazi:2006wu}
M.~S. Alenazi and P.~Gondolo,
\newblock Phys. Rev. D {\bf 74}, 083518 (2006), astro-ph/0608390.

\bibitem{Collar:1992qc}
J.~I. Collar and F.~T. Avignone,
\newblock Phys. Lett. B {\bf 275}, 181 (1992).

\bibitem{Collar:1993ss}
J.~I. Collar and F.~T. Avignone, III,
\newblock Phys. Rev. D {\bf 47}, 5238 (1993).

\bibitem{Hasenbalg:1997hs}
F.~Hasenbalg {\em et~al.},
\newblock Phys. Rev. D {\bf 55}, 7350 (1997), astro-ph/9702165.

\bibitem{damadaily}
R.~Bernabei {\em et~al.},
\newblock The European Physical Journal C {\bf 74}, 2827 (2014).

\bibitem{luxdaily}
LUX Collaboration, D.~S. Akerib {\em et~al.},
\newblock Phys. Rev. D {\bf 98}, 062005 (2018).

\bibitem{pandadaily}
PandaX-II Collaboration, X.~Cui {\em et~al.},
\newblock Phys. Rev. Lett. {\bf 128}, 171801 (2022).

\bibitem{prospectdaily}
PROSPECT Collaboration, M.~Andriamirado {\em et~al.},
\newblock Phys. Rev. D {\bf 104}, 012009 (2021).

\bibitem{Holdom1986}
B.~{Holdom},
\newblock Physics Letters B {\bf 166}, 196 (1986).

\bibitem{1984PhLB..136..279G}
P.~{Galison} and A.~{Manohar},
\newblock Physics Letters B {\bf 136}, 279 (1984).

\bibitem{Emken:2019tni}
T.~Emken, R.~Essig, C.~Kouvaris, and M.~Sholapurkar,
\newblock JCAP {\bf 09}, 070 (2019), 1905.06348.

\bibitem{Kouvaris:2014lpa}
C.~Kouvaris and I.~M. Shoemaker,
\newblock Phys. Rev. D {\bf 90}, 095011 (2014), 1405.1729.

\bibitem{Avalos_2021}
N.~Ávalos {\em et~al.},
\newblock Journal of Physics: Conference Series {\bf 2156}, 012074 (2021).

\bibitem{damic_idm2022}
I.~Arnquist {\em et~al.},
\newblock {T}he {DAMIC-M} {E}xperiment: {S}tatus and {F}irst {R}esults, 2022,
  arXiv:2210.12070.

\bibitem{Holland:2002zz}
S.~E. Holland,
\newblock Exper. Astron. {\bf 14}, 83 (2002).

\bibitem{Holland:2003zz}
S.~E. Holland, D.~E. Groom, N.~P. Palaio, R.~J. Stover, and M.~Wei,
\newblock IEEE Transactions on Electron Devices {\bf 50}, 225 (2003).

\bibitem{Holland:2009zz}
S.~E. Holland, W.~F. Kolbe, and C.~J. Bebek,
\newblock IEEE Transactions on Electron Devices {\bf 56}, 2612 (2009).

\bibitem{ALESSANDRELLO1991106}
A.~Alessandrello {\em et~al.},
\newblock Nuclear Instruments and Methods in Physics Research Section B: Beam
  Interactions with Materials and Atoms {\bf 61}, 106 (1991).

\bibitem{geant4}
S.~Agostinelli {\em et~al.},
\newblock Nuclear Instruments and Methods in Physics Research Section A:
  Accelerators, Spectrometers, Detectors and Associated Equipment {\bf 506},
  250 (2003).

\bibitem{PhysRevD.94.082006}
DAMIC Collaboration, A.~Aguilar-Arevalo {\em et~al.},
\newblock Phys. Rev. D {\bf 94}, 082006 (2016).

\bibitem{skipper}
J.~Janesick {\em et~al.},
\newblock Proc. SPIE {\bf 1242}, 223 (1990).

\bibitem{Chandler1990zz}
C.~E. Chandler {\em et~al.},
\newblock Proc. SPIE {\bf 1242}, 238 (1990).

\bibitem{Tiffenberg:2017aac}
J.~Tiffenberg {\em et~al.},
\newblock Phys. Rev. Lett. {\bf 119}, 131802 (2017).

\bibitem{Compton-DAMICM}
DAMIC-M Collaboration, D.~Norcini {\em et~al.},
\newblock Phys. Rev. D {\bf 106}, 092001 (2022).

\bibitem{2001sccd.book.....J}
J.~R. {Janesick},
\newblock {\em {Scientific charge-coupled devices}} (SPIE Optical Engineering
  Press, 2001).

\bibitem{Note1}
The time of each image was recorded in Coordinated Universal Time (UTC).

\bibitem{Cowan:2010js}
G.~Cowan, K.~Cranmer, E.~Gross, and O.~Vitells,
\newblock Eur. Phys. J. C {\bf 71}, 1554 (2011), 1007.1727,
\newblock [Erratum: Eur.Phys.J.C 73, 2501 (2013)].

\bibitem{Gross:2010qma}
E.~Gross and O.~Vitells,
\newblock Eur. Phys. J. C {\bf 70}, 525 (2010), 1005.1891.

\bibitem{Lee:2015qva}
S.~K. Lee, M.~Lisanti, S.~Mishra-Sharma, and B.~R. Safdi,
\newblock Phys. Rev. D {\bf 92}, 083517 (2015), 1508.07361.

\bibitem{Griffin:2021znd}
S.~M. Griffin, K.~Inzani, T.~Trickle, Z.~Zhang, and K.~M. Zurek,
\newblock Phys. Rev. D {\bf 104}, 095015 (2021), 2105.05253.

\bibitem{Green:2017odb}
A.~M. Green,
\newblock J. Phys. G {\bf 44}, 084001 (2017), 1703.10102.

\bibitem{Note2}
We use the Standard Halo Model~\cite {PhystatDM} parameters with the Earth's
  velocity fixed to its average value during the data-taking period,
  $v_{E}=263$~km/s.

\bibitem{McCabe:2013kea}
C.~McCabe,
\newblock JCAP {\bf 02}, 027 (2014), 1312.1355.

\bibitem{Emken:2017qmp}
T.~Emken and C.~Kouvaris,
\newblock JCAP {\bf 10}, 031 (2017), 1706.02249.

\bibitem{Starkman:212913}
G.~D. Starkman, A.~Gould, R.~Esmailzadeh, and S.~K. Dimopoulos,
\newblock Phys. Rev. D {\bf 41}, 3594 (1990).

\bibitem{Kavanagh:2016pyr}
B.~J. Kavanagh, R.~Catena, and C.~Kouvaris,
\newblock JCAP {\bf 01}, 012 (2017), 1611.05453.

\bibitem{Mahdawi:2017cxz}
M.~S. Mahdawi and G.~R. Farrar,
\newblock JCAP {\bf 12}, 004 (2017), 1709.00430.

\bibitem{Hooper:2018bfw}
D.~Hooper and S.~D. McDermott,
\newblock Phys. Rev. D {\bf 97}, 115006 (2018), 1802.03025.

\bibitem{Mahdawi:2018euy}
M.~S. Mahdawi and G.~R. Farrar,
\newblock JCAP {\bf 10}, 007 (2018), 1804.03073.

\bibitem{Kavanagh:2017cru}
B.~J. Kavanagh,
\newblock Phys. Rev. D {\bf 97}, 123013 (2018), 1712.04901.

\bibitem{Verne}
B.~J. Kavanagh,
\newblock {Verne [Code, v1.3]},
\newblock \url{https://github.com/bradkav/verne},
  \href{https://doi.org/10.5281/zenodo.7193430}{DOI:10.5281/zenodo.7193430},
  2017-2023.

\bibitem{Note3}
Note that \protect \textsc {Verne}\protect \xspace uses an alternative
  convention for the isodetection angle $\gamma = 180^\circ - \Theta $.

\bibitem{Lantero_inprep}
A.~Lantero-Barreda and B.~J. Kavanagh,
\newblock in preparation, 2024.

\bibitem{Note4}
Note that the isodetection angle corresponding to a given time in a sidereal
  day changes by several degrees during the data taking period.

\bibitem{SENSEIrecast}
The recast limit, provided by the SENSEI Collaboration, uses the same halo
  parameters and charge yield model of Ref.~\cite{DAMIC-M:2023gxo} for proper
  comparison .

\bibitem{xenon1t2019}
XENON Collaboration, E.~Aprile {\em et~al.},
\newblock Phys. Rev. Lett. {\bf 123}, 251801 (2019).

\bibitem{xenon1t-SE-2022}
XENON Collaboration, E.~Aprile {\em et~al.},
\newblock Phys. Rev. D {\bf 106}, 022001 (2022).

\bibitem{PandaX-2021}
PandaX-II Collaboration, C.~Cheng {\em et~al.},
\newblock Phys. Rev. Lett. {\bf 126}, 211803 (2021).

\bibitem{xenon10_2017}
R.~Essig, T.~Volansky, and T.-T. Yu,
\newblock Phys. Rev. D {\bf 96}, 043017 (2017).

\bibitem{SolarReflectedDM}
H.~An, H.~Nie, M.~Pospelov, J.~Pradler, and A.~Ritz,
\newblock Phys. Rev. D {\bf 104}, 103026 (2021).

\bibitem{PhysRevD.105.063020}
T.~Emken,
\newblock Phys. Rev. D {\bf 105}, 063020 (2022), 2102.12483.

\bibitem{USCosmicVisions2017}
M.~Battaglieri {\em et~al.},
\newblock {US} {C}osmic {V}isions: {N}ew {I}deas in {D}ark {M}atter 2017:
  {C}ommunity {R}eport, arXiv:1707.04591.

\bibitem{PhysRevD.102.063026}
K.~Ramanathan and N.~Kurinsky,
\newblock Phys. Rev. D {\bf 102}, 063026 (2020).

\bibitem{PhysRevD.105.062003}
DAMIC Collaboration, A.~Aguilar-Arevalo {\em et~al.},
\newblock Phys. Rev. D {\bf 105}, 062003 (2022).

\bibitem{asymptoticlikelihood}
G.~Cowan, K.~Cranmer, E.~Gross, and O.~Vitells,
\newblock The European Physical Journal C {\bf 71}, 1554 (2011).

\bibitem{Note5}
Complementary searches for semi-relativistic DM fluxes from Solar
  reflection~\cite {SolarReflectedDM,PhysRevD.105.063020}, and
  cosmological~\cite {Krnjaic:2019dzc,Giovanetti:2021izc} and stellar evolution
  constraints~\cite {Davidson:2000hf,Chang:2018rso} also place limits in this
  mass range.

\bibitem{PhystatDM}
D.~Baxter {\em et~al.},
\newblock The European Physical Journal C {\bf 81}, 907 (2021).

\bibitem{Krnjaic:2019dzc}
G.~Krnjaic and S.~D. McDermott,
\newblock Phys. Rev. D {\bf 101}, 123022 (2020), 1908.00007.

\bibitem{Giovanetti:2021izc}
C.~Giovanetti, M.~Lisanti, H.~Liu, and J.~T. Ruderman,
\newblock Phys. Rev. Lett. {\bf 129}, 021302 (2022), 2109.03246.

\bibitem{Davidson:2000hf}
S.~Davidson, S.~Hannestad, and G.~Raffelt,
\newblock JHEP {\bf 05}, 003 (2000), hep-ph/0001179.

\bibitem{Chang:2018rso}
J.~H. Chang, R.~Essig, and S.~D. McDermott,
\newblock JHEP {\bf 09}, 051 (2018), 1803.00993.

\bibitem{cdex10}
CDEX Collaboration, Z.~Y. Zhang {\em et~al.},
\newblock Phys. Rev. Lett. {\bf 129}, 221301 (2022).

\bibitem{PhysRevLett.130.101002}
DarkSide Collaboration, P.~Agnes {\em et~al.},
\newblock Phys. Rev. Lett. {\bf 130}, 101002 (2023).

\bibitem{Knapen_2022}
S.~Knapen, J.~Kozaczuk, and T.~Lin,
\newblock Phys. Rev. D {\bf 105}, 015014 (2022).

\bibitem{Griffin_2021}
S.~M. Griffin, K.~Inzani, T.~Trickle, Z.~Zhang, and K.~M. Zurek,
\newblock Phys. Rev. D {\bf 104}, 095015 (2021).

\bibitem{exceedDMpublished}
T.~Trickle,
\newblock Phys. Rev. D {\bf 107}, 035035 (2023).

\bibitem{Emken:2018run}
T.~Emken and C.~Kouvaris,
\newblock Phys. Rev. D {\bf 97}, 115047 (2018), 1802.04764.

\bibitem{DaMaSCUS}
T.~Emken and C.~Kouvaris,
\newblock {Dark Matter Simulation Code for Underground Scatterings~(DaMaSCUS)
  [Code, v1.1]},
\newblock \url{https://github.com/temken/damascus},
  \href{https://doi.org/10.5281/zenodo.3726878}{DOI:10.5281/zenodo.3726878},
  2017-2020.

\end{thebibliography}

\clearpage
\newpage
\section*{Supplemental Material}

Independent cross-checks have been performed at every step in the analysis, starting from the low-level image processing to the generation of the data pixel distribution, the identification of defects, the modeling of the DM signal, and the extraction of the DM signal upper limit. All of these checks yield consistent results. 

\subsection*{Rates}
The observed time dependence of the rate $R_{1}^i$ is consistent with  dark current stabilizing over time, characteristic in these types of devices. To confirm this interpretation we check the time dependence of the 2$e^-$ rate, $R_{2}^i$, in the same data set. We sum the pixel charge distribution of 100 consecutive images to guarantee a sufficient number of pixels with 2$e^-$ and then fit it with the sum of 3 Gaussians to obtain $R_{2}^i$. The exponential time decay constant for $R_{2}^i$, $\tau_{2e}=35 \pm 24$~days, when compared with that for $R_{1}^i$ is consistent with a Poisson process, as expected for dark current. 

\subsection*{Model-independent search}
A deviation from the null hypothesis (no modulation) is observed for T=26\,h in the L-side. The best fit amplitude when fixing T=26\,h is found to be $A = (594 \pm 103)$\,events/g/day  with a phase $\phi = (10.5\pm 0.7)$\,h. The best fit amplitude for T=24\,h is found to be $A = (229 \pm 103)$\,events/g/day with a phase $\phi = (6.7\pm 1.7)$\,h. This cannot be interpreted as evidence for a DM signal, not only because it does not match the expected period, but also because a much higher deviation should have been found in the U-side, which accounts for 70\% of the data. In fact, for the U-side, the best fit point for amplitudes and phases is $A = (141 \pm 42)$\,events/g/day and $\phi = (7.7\pm 1.2)$\,h when T=26\,h. For T=24\,h, instead, $A = (121 \pm 42)$\,events/g/day and $\phi = (16.8\pm 1.3)$\,h.
The same argument is valid for changes in external parameters like the temperature, which should affect the measured rate in both the U and L-side. Also, no modulation is found in relevant slow control data (several temperature sensors in the apparatus and cryocooler parameters). 
Applying the same model-independent search on the data set previously excluded due to the presence of serial register traps (which should have no effect on a possible modulation signal), no evidence for a T=26\,h signal is found.
Note that the L-side data consists of pixels close to the amplifier (columns~$>74$ are rejected), and the modulation amplitude is reduced by a factor of two when all columns are included. From these observations, an instrumental effect limited to the amplifier on the L side is likely the origin of the deviation. Note that since we took data for over two months a T=26\,h modulation is washed out when searching for a signal at a sidereal period (23.93\,h), hence we do not expect any significant effect on the DM-$e^{-}$ scattering analysis. Indeed, excluding the L-side data changes the exclusion limits in Fig.~\ref{fig:upperlimits} by at most 5\% (12\%) for the ultralight (heavy) mediator. As our primary objective is to assess the temporal stability of the detector by quantifying the statistical significances of fluctuations in the data, the presence of a modulated signal highlights the complexity of the detector's time dependence. We acknowledge the need for a comprehensive understanding of these temporal variations. To address this, the method outlined involves a comparison of best-fit amplitudes and phases across different amplifiers or CCDs. This cross-check enables us to discern any signal origin that might be common to all CCDs, aiding in the ongoing effort to refine our understanding of the detector's time-dependent behavior and ensuring the robustness of our results against potential instrumental artifacts in future searches.

\subsection*{Exclusion limits}
For the model-independent analysis, we have performed toy-MC simulations under $H_0$ (no modulation), finding that the test statistic $t_q$  asymptotically follows a $\chi^2$ distribution with two degrees of freedom. 

We have also verified that our limits in Fig.~\ref{fig:upperlimits} have the correct coverage by performing the analysis on toy Monte Carlo simulations, which include both a signal and background. We find that the signal is properly rejected or discovered according to our 90\% C.L. 

\begin{figure*}
    \centering
	\hspace{-0.2cm}
        \includegraphics[width=0.48\textwidth]{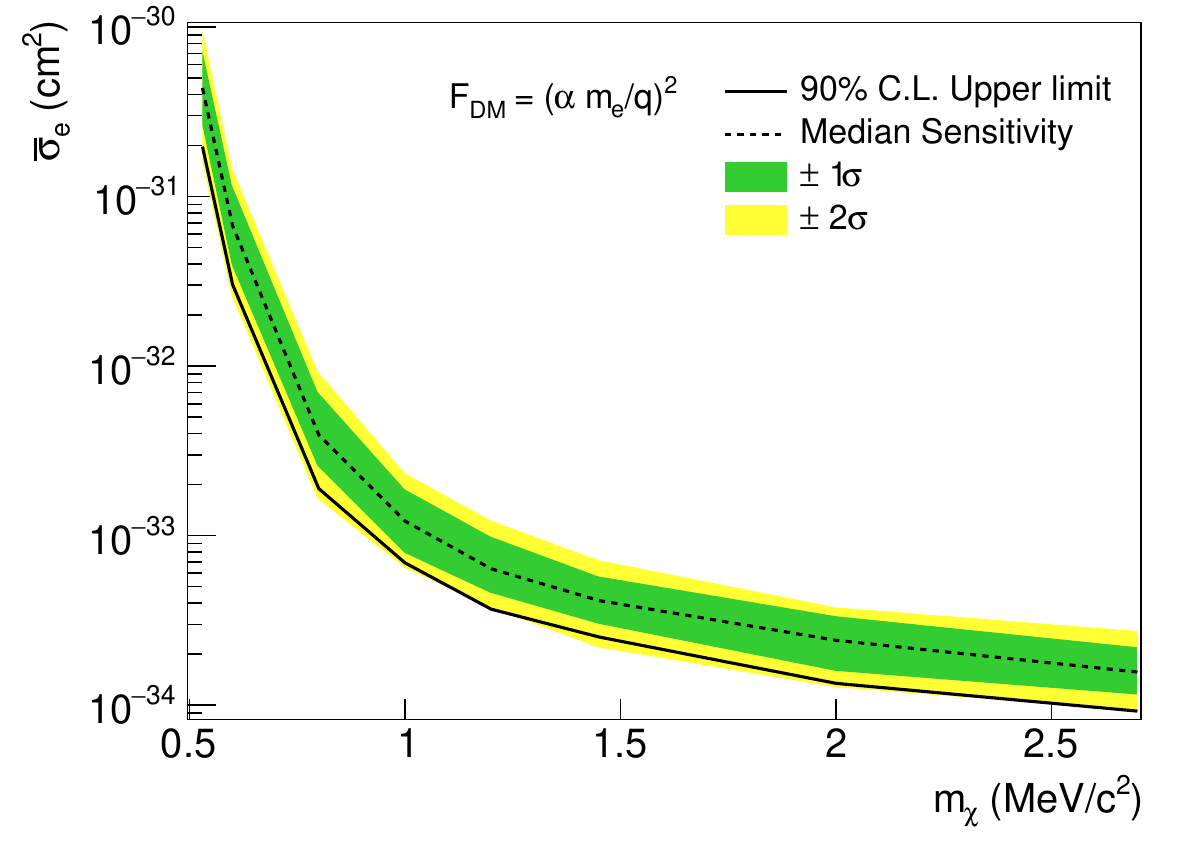}
        \hspace{0.6cm}
   	\includegraphics[width=0.48\textwidth]{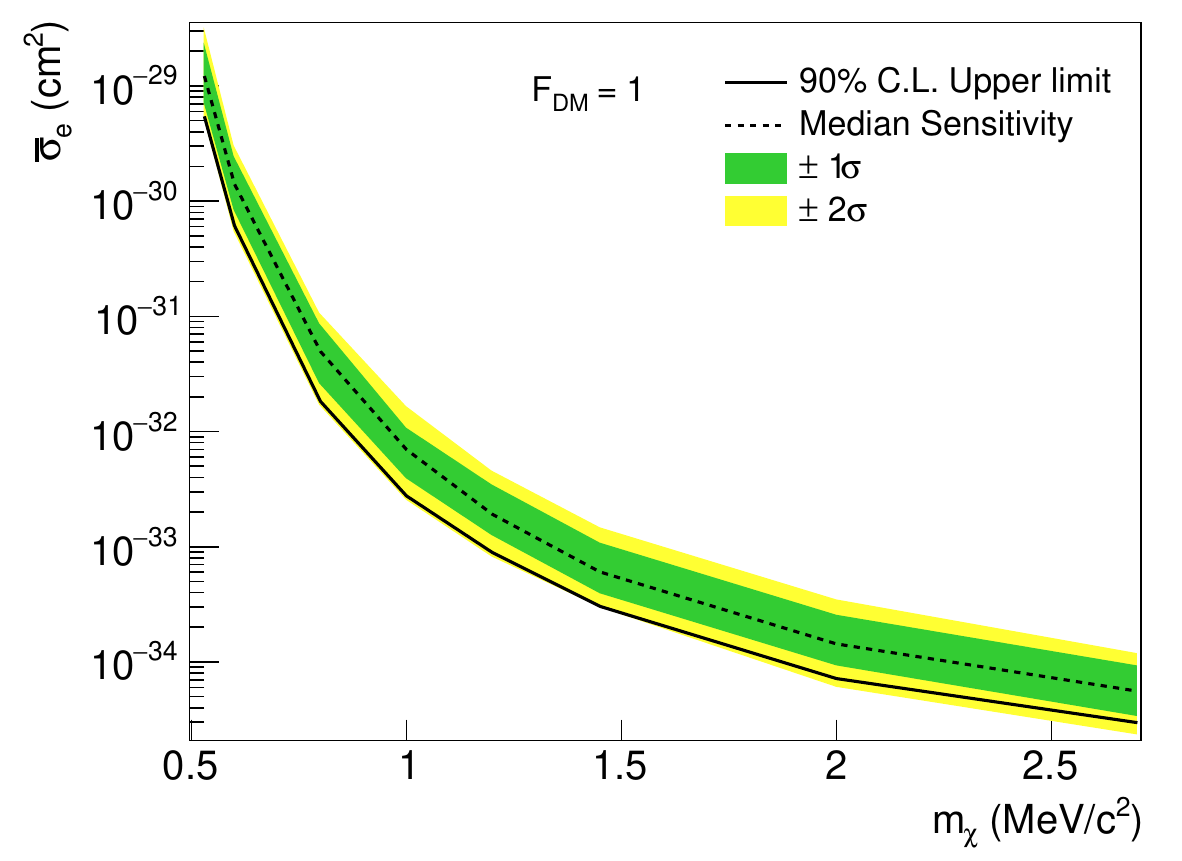}
    \caption{Sensitivity bands for DM-electron interactions via a ultralight (left) and heavy (right) mediator. Also shown is the median sensitivity (dashed line) and the DAMIC-M upper limit from Fig.~\ref{fig:upperlimits} (solid line).}
\label{fig:sensitivity}
\end{figure*}

In Fig.~\ref{fig:sensitivity} we show the estimated sensitivity bands. The bands are obtained from 10000 pseudo data sets constructed by randomly shuffling the time of the images. The shuffling method guarantees that no modulation is present in the pseudo data set and takes into account the uncertainty in the measured rates. A $90\%$ C.L upper limit is then calculated for each pseudo data set, and sensitivity bands are derived from the distribution of these upper limits. The limit reported in Fig.~\ref{fig:upperlimits} falls in the $2\sigma$ (95.4$\%$) sensitivity band.

We also perform a cross-check of the exclusion limits with a less sensitive but simpler method by computing the correlation 
\begin{equation}    
\rho = \frac{\sqrt{2}}{N_{\rm{im}}||R_1||}\sum_{i=1}^{N_{\rm{im}}} R_1^i \cos\left[\frac{2\pi(t_i-\phi)}{T}\right]\,,
\end{equation}
where $||R_1||=\sqrt{\sum_{i=1}^{N_{\rm{im}}}(R_1^i)^2/N_{\rm{im}}} $
and $T$ and $\phi$ are fixed to the period and phase predicted by the DM-$e^-$ scattering model described in the main text. The probability distribution of $\rho$ is obtained from toy Monte Carlo simulations, which include only the background model, and then used to derive a 90\% limit on the amplitude of a putative daily modulation signal. We convert the rate to cross section using Eq.~\eqref{eq:dR_dEe}. The corresponding exclusion limits are a factor of 1.2 weaker than the nominal limits obtained from the likelihood.

\begin{figure*}
    \centering
	\hspace{-0.2cm}
        \includegraphics[width=0.48\textwidth]{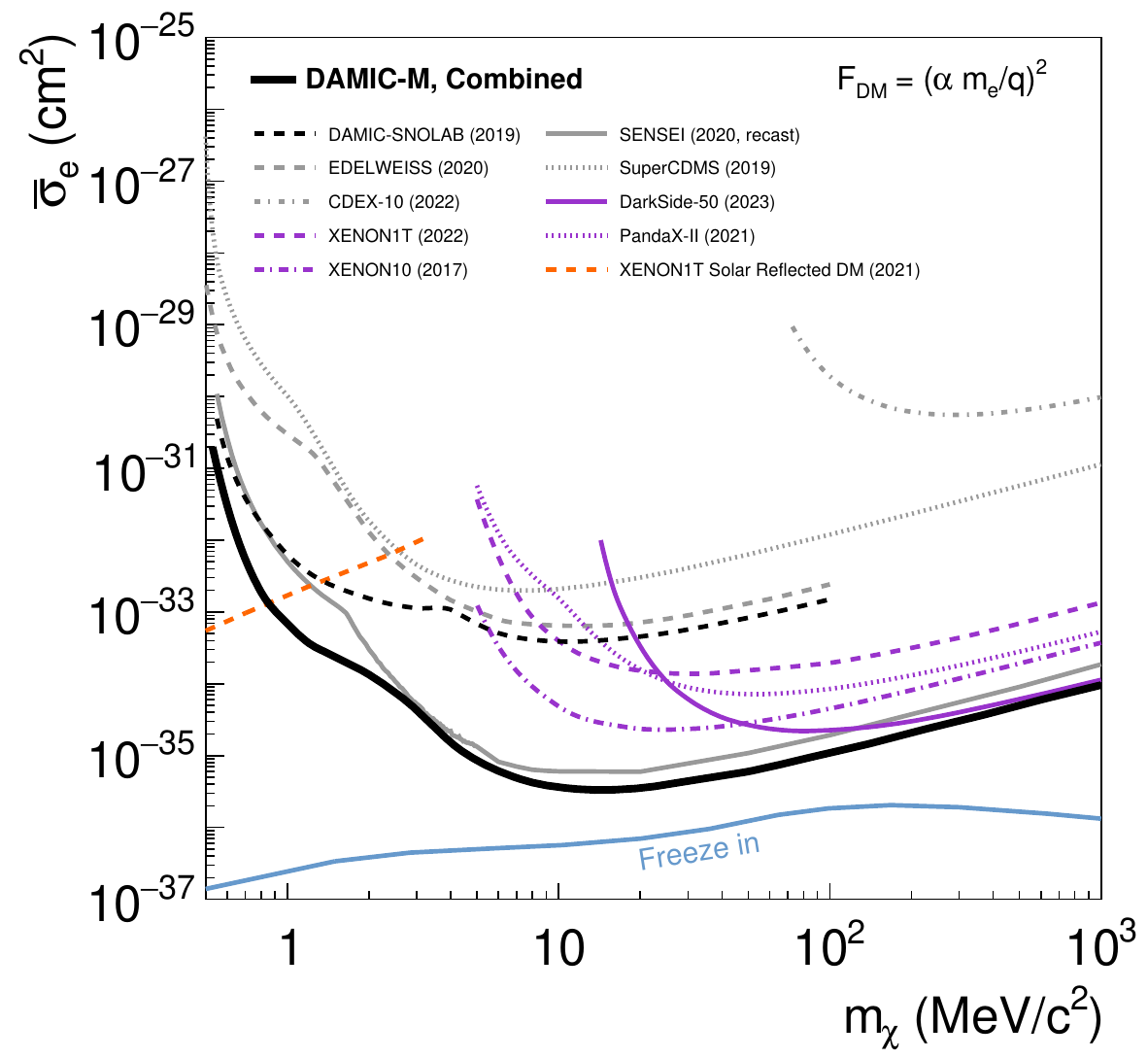}
        \hspace{0.6cm}
   	\includegraphics[width=0.48\textwidth]{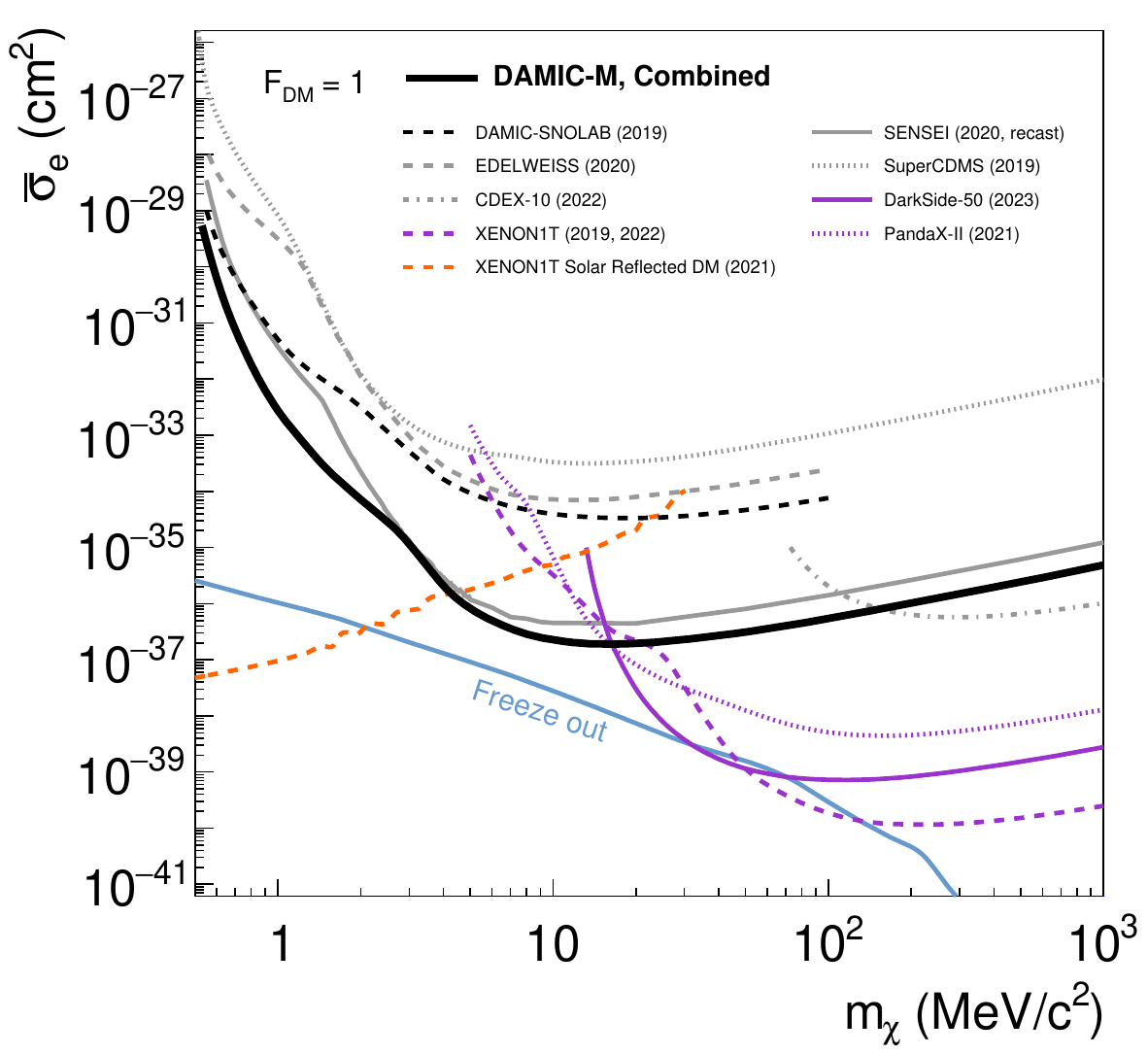}
    \caption{DAMIC-M 90\% C.L. upper limits (solid thick black) on DM-electron interactions through an ultralight (left) and heavy (right) dark photon mediator obtained combining the daily modulation analysis with the previous DAMIC-M result~\cite{DAMIC-M:2023gxo}. Also shown are previous limits from other experiments: DAMIC-SNOLAB~\cite{DAMIC:2019dcn} (dashed black); SENSEI~\cite{SENSEI:2020dpa,SENSEIrecast} (solid gray); EDELWEISS~\cite{EDELWEISS:2020fxc} (dashed gray); SuperCDMS~\cite{SuperCDMS:2018mne} (dotted gray);
    CDEX-10\, \cite{cdex10} (dot-dashed gray); DarkSide-50\,\cite{PhysRevLett.130.101002} (solid violet); 
    XENON1T combined result from~\cite{xenon1t2019,xenon1t-SE-2022} (dashed violet); PandaX-II~\cite{PandaX-2021} (dotted violet); a limit obtained from XENON10 data in Ref.~\cite{xenon10_2017} (dash-dotted violet); and a limit obtained from XENON1T data considering ``solar reflected DM" (dashed orange) from Ref.~\cite{SolarReflectedDM} (left) and  Ref.~\cite{PhysRevD.105.063020} (right). Theoretical expectations assuming a DM relic abundance from freeze-in and freeze-out mechanisms are also shown in light blue~\cite{USCosmicVisions2017}. 
    }
\label{fig:combinedlimit}
 \end{figure*}

In Fig.~\ref{fig:combinedlimit} we present DAMIC-M 90\% C.L. exclusion limits for ultralight (left) and heavy (right) mediators up to masses of 1000~MeV/c$^2$, obtained by combining results in Fig.~\ref{fig:upperlimits} with our previous limits~\cite{DAMIC-M:2023gxo}.
 
\subsection*{Theoretical uncertainties}
We use $\texttt{QEDark}$ as the reference theoretical model for proper comparison with previous and forthcoming results from other experiments. We note that theoretical uncertainties are significant for DM masses below $2.7$\,MeV/c$^2$, with
$\texttt{QEDark}$-based limits almost a factor 100 better than those obtained with $\texttt{DarkELF}$~\cite{Knapen_2022} and $\texttt{EXCEED-DM}$~\cite{Griffin_2021,exceedDMpublished} models (see Supplemental Material of Ref.~\cite{DAMIC-M:2023gxo}.)

\subsection*{Earth-scattering calculations}

The velocity distribution at the detector was determined with an updated version of the \Verne code to take into account Earth-scattering effects. This code will be described in full in a future publication (Ref.~\cite{Lantero_inprep}). Here we provide a brief summary of the calculation, including signal rates and modulation amplitude for a number of benchmark parameter points, and a comparison with other Earth-scattering calculations available in the literature.

The original \Verne code~\cite{Verne}, described in Ref.~\cite{Kavanagh:2017cru}, was tailored to the Earth-scattering of heavy DM particles. It assumed straight-line trajectories for the incoming DM particles and a continuous energy loss due to scattering with nuclei in the atmosphere and Earth. Instead, light DM ($m_\chi \ll m_N$) may be deflected substantially when scattering off a nucleus of mass $m_N$, while the energy losses in a single scattering event (proportional to $m_\chi/m_N$) are typically negligible. The updated version of \Verne used in the present work assumes that the DM particles follow a straight-line trajectory until they scatter, at which point they either continue along the same trajectory, or they are reflected back along the incoming trajectory with probability $p_\mathrm{back}$. The value of $p_\mathrm{back}$ depends on the type of DM interaction and a suitable estimate can be calculated as the fraction of scattering events which deflect the DM particle into the backwards hemisphere. We fix $p_\mathrm{back} = 0.875$ and $p_\mathrm{back} = 0.5$ for the light and heavy mediator cases respectively. We assume zero energy loss with each scattering and take into account up to two scattering events for each incoming trajectory. The flux at the detector is given by the sum of those particles which arrive at the detector without being reflected out to space and those particles which are reflected back to the detector, having already passed the detector. Full details of the formalism and implementation will appear in Ref.~\cite{Lantero_inprep}.

\begin{figure}[H]
    \centering
    \includegraphics[width=0.48\textwidth]{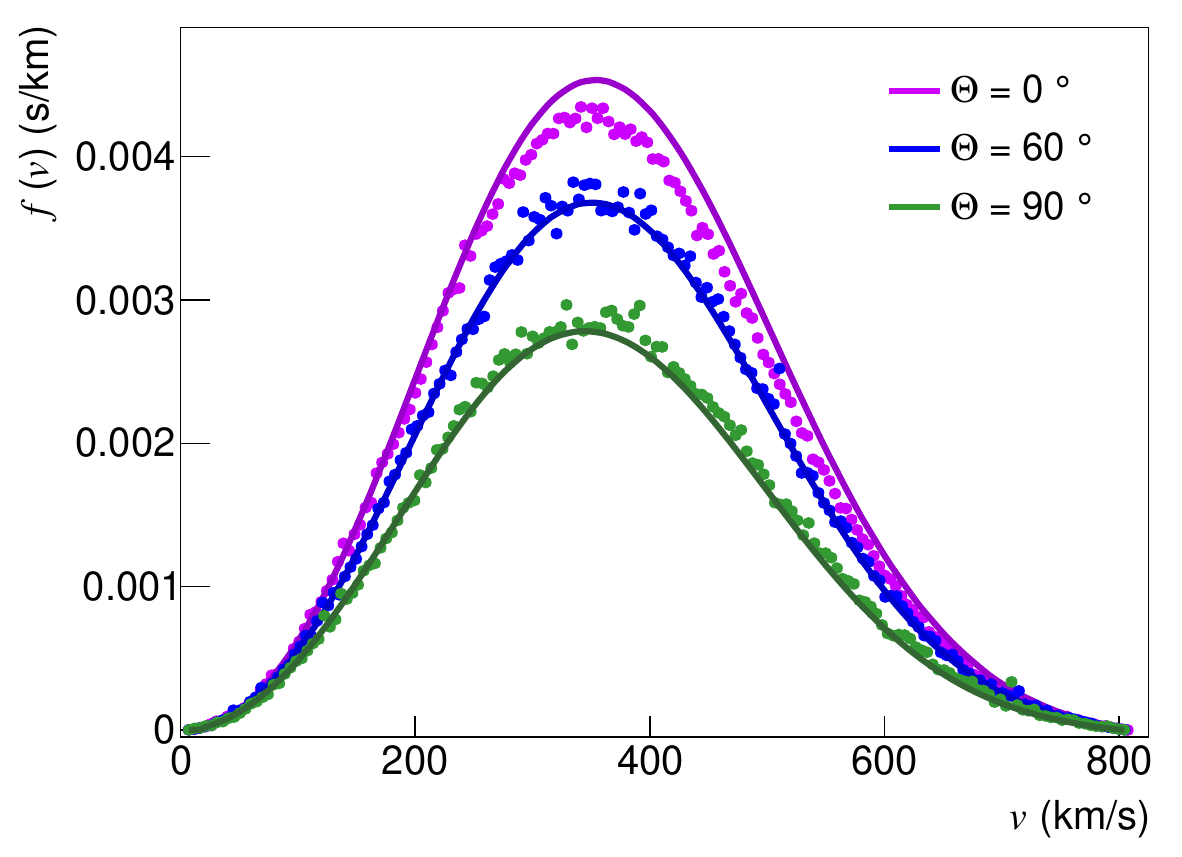}
    \caption{ Speed distributions at different isodetection angles for DM particles of mass 1\,MeV/c$^2$ interacting via an ultralight dark photon mediator (\Verne, solid line; \DaMaSCUS, dots). The distributions are calculated for $\bar{\sigma}_e=10^{-33}$\,cm$^2$, close to the DAMIC-M limit at m$_\chi$=1\,MeV/c$^2$ reported in this letter. }
    \label{fig:VerneDamascus}
\end{figure}

In Fig.~\ref{fig:VerneDamascus} we compare the \Verne DM velocity distributions at the detector with those obtained when the Earth-scattering effects are calculated using \DaMaSCUS~\cite{Emken:2017qmp,Emken:2018run,DaMaSCUS}, a more accurate but computationally demanding 3D Monte Carlo simulation. The simplified approach adopted in \Verne reproduces to better than 20\% the \DaMaSCUS predictions. 
We have repeated our analysis at a DM mass of 1 MeV/c$^2$ using the \DaMaSCUS code; the corresponding limits differ from those obtained with \Verne by less than 30\%. The use of \Verne has also allowed for a number of additional checks. In the \Verne calculation of the DM velocity distribution the Earth's speed is fixed to $263$~km/s. To check the effect of this assumption we repeat the analysis using the largest and smallest value of $\mathrm{v}_\mathrm{E}$, $267$ and $258$~km/s, corresponding to the beginning and the end of the data-taking period. The exclusion limits change by at most 15\%. 

For illustration, we report in Table~\ref{tab:rates_lDP} and ~\ref{tab:rates_hDP} the mean DM-electron signal rate $\langle R \rangle$ and the amplitude $A$ of the daily modulation, calculated using \Verne and \DaMaSCUS. The values are obtained for representative DM masses $m_\chi$, and for cross sections $\bar{\sigma}_e$ close to the upper limits reported in Fig.~\ref{fig:upperlimits}. We find that the maximum difference between the two calculations is $32\%$ for the modulation amplitude and $5\%$ for the mean rate.

\begin{table*}
\centering
\begin{tabular}{cccccc}
\hline
$m_\chi$ (MeV)& $\bar{\sigma}_e$ [cm$^2$] & $A_{D}$ ($\frac{\mathrm{events}}{\mathrm{g}\cdot \mathrm{day}}$) & $A_{V}$ ($\frac{\mathrm{events}}{\mathrm{g}\cdot \mathrm{day}}$) &  $\langle R_{D} \rangle$ ($\frac{\mathrm{events}}{\mathrm{g}\cdot \mathrm{day}}$) & $\langle R_{V} \rangle$ ($\frac{\mathrm{events}}{\mathrm{g}\cdot \mathrm{day}}$) \\
\hline
0.53 & \( 10^{-31}\) & 20 & 20 & 34 & 34 \\
1.0 & \( 10^{-33}\) & 45 & 51 & 150 & 157 \\
2.7 & \( 10^{-34}\) & 22 & 29 & 185 & 178 \\
\hline
\end{tabular}

\caption{Modulation amplitude $A$ and mean DM-electron signal rate $\langle R \rangle$ calculated with VERNE (V) and DaMaSCUS (D) for the ultralight mediator model. The mean $\langle R \rangle$ is obtained by averaging the rate over time.  The modulation amplitude is defined as the maximum deviation from the mean rate, $A = \mathrm{max}(R^{max}-\langle R \rangle, \langle R \rangle-R^{min})$, for the first day of data taking. We use this amplitude definition because the daily modulation is not symmetric with respect to the mean rate.}
\label{tab:rates_lDP}

\vspace{5em}

\centering
\begin{tabular}{cccccc}
\hline
$m_\chi$ (MeV)& $\bar{\sigma}_e$ [cm$^2$] & $A_{D}$ ($\frac{\mathrm{events}}{\mathrm{g}\cdot \mathrm{day}}$) & $A_{V}$ ($\frac{\mathrm{events}}{\mathrm{g}\cdot \mathrm{day}}$) &  $\langle R_{D} \rangle$ ($\frac{\mathrm{events}}{\mathrm{g}\cdot \mathrm{day}}$) & $\langle R_{V} \rangle$ ($\frac{\mathrm{events}}{\mathrm{g}\cdot \mathrm{day}}$) \\
\hline
0.53 & \( 10^{-30}\) & 6 & 6 & 10 & 11 \\
1.0 & \( 10^{-32}\) & 135 & 130 & 278 & 273 \\
2.7 & \( 10^{-35}\) & 6 & 7 & 25 & 26 \\
\hline
\end{tabular}

\caption{
The explanation can be found in Tab.~\ref{tab:rates_lDP} 
}
\label{tab:rates_hDP}
\end{table*}

\end{document}